\documentclass[floatfix, aps, apl, superscriptaddress, reprint]{revtex4-1}

\usepackage{graphicx}               
\usepackage{bm}                     
\usepackage{amsfonts,amssymb}       
\usepackage{dcolumn}                
\newcommand*\pct{\scalebox{.9}{\%}}

\begin{document}

\title{Hinge-like structure induced unusual properties of black phosphorus and new strategies to improve the thermoelectric performance}

\author{Guangzhao~Qin}
\author{Qing-Bo~Yan}
\email{yan@ucas.ac.cn}
\affiliation{College of Materials Science and Opto-Electronic Technology, University of Chinese Academy of Sciences, Beijing 100049, People's Republic of China}
\author{Zhenzhen~Qin}
\affiliation{College of Electronic Information and Optical Engineering, Nankai University, Tianjin 300071, People's Republic of China}
\author{Sheng-Ying~Yue}
\author{Hui-Juan~Cui}
\author{Qing-Rong~Zheng}
\author{Gang~Su}
\email{gsu@ucas.ac.cn}
\homepage{http://tcmp2.ucas.ac.cn/}
\affiliation{School of Physics, University of Chinese Academy of Sciences, Beijing 100049, People's Republic of China}

\date{\today}

\begin{abstract}
We systematically investigated the geometric, electronic and thermoelectric (TE)
properties of bulk black phosphorus (BP) under strain. The hinge-like structure of BP brings unusual mechanical responses such as anisotropic
Young's modulus and negative Poisson's ratio. A sensitive electronic structure of BP makes it transform among metal, direct and indirect semiconductors under strain.
The maximal figure of merit \emph{ZT} of BP is found to be 0.72 at $800\,\mathrm{K}$ that could be enhanced to 0.87 by exerting an appropriate strain,
revealing BP could be a potential medium-high temperature TE material. Such strain-induced enhancements of TE performance are often observed to occur at the boundary of the
direct-indirect band gap transition, which can be attributed to the increase of degeneracy of energy valleys at the transition point.
By comparing the structure of BP with SnSe, a family of potential TE materials with hinge-like structure are suggested. This study not only exposes various novel properties of BP under strain,
but also proposes effective strategies to seek for better TE materials.

\end{abstract}

\pacs{}
\maketitle


Thermoelectric (TE) materials can perform a direct solid-state conversion from thermal
to electrical energy or vice versa, and have a number of valuable applications,
such as thermoelectric generators, waste heat recovery, thermoelectric cooling and heating devices \cite{C3TA14259K, Sales15022002, Bell12092008, MRS:7960292}, etc., which thus may make crucial contributions to the crisis of energy and environment.
In general, the TE performance and efficiency are characterized by the
dimensionless figure of merit $ZT = S^2\sigma T/\kappa$, where $S$, $\sigma$,
$T$ and $\kappa$ are Seebeck coefficient (thermopower), electrical conductivity,
absolute temperature and thermal conductivity, respectively.
The thermal conductivity ($\kappa = \kappa_e + \kappa_{ph}$) consists of those from
electrons ($\kappa_e$) and phonons
($\kappa_{ph}$). \cite{SnyderTobererNatureThermoelectric}
Accordingly, a higher \emph{ZT} value at a given temperature requires a high thermopower, a suitable
combination of electrical conductivity and electrical thermal conductivity (both
related to carrier mobility), and a low lattice thermal conductivity.

Among the known TE materials, Bi$_2$Te$_3$ and Bi$_2$Se$_3$ perform the best at room temperature. They are narrow band-gap
semiconductors with layered structures, and have a high electrical conductivity in
combination with a low thermal conductivity, resulting in a high \emph{ZT} value.
\cite{0508-3443-9-9-306, SECTOBNature413597}
Recently, another layered material, SnSe, has been reported to have an
unprecedented high \emph{ZT} value of 2.6 at $930\,\mathrm{K}$ along a specific
lattice direction.\cite{ZLLSZYSHTGUCWCDK} A closer inspection indicates that
both Bi$_2$Te$_3$ (Bi$_2$Se$_3$) and SnSe share the similar layered structures. It thus
reminds us that a high \emph{ZT} value along some specific lattice directions may
emerge in the layered anisotropic materials.
Such a speculation motivates us to reexamine the TE performance of black
phosphorus (BP) that has an almost the same hexagonal layered structure as SnSe
except the lowered symmetry of SnSe crystal (Supplemental Fig.~S1), because
for each layer, BP and SnSe bear a nearly identical hinge-like structure, which may
lead to a high electrical conductivity and a low lattice thermal conductivity.  \cite{ZLLSZYSHTGUCWCDK}
BP was also found to be a direct band gap semiconductor with a high carrier
mobility \cite{JPSJ.52.2148, FIELD_EFFECT_BP}. Besides,
Few-layer BP (the single layer BP is called phosphorene),
which possesses the layered hexagonal structure similar to graphene but puckered,
has been successfully mechanically exfoliated recently \cite{ACSNano84033nn501226z, FIELD_EFFECT_BP},
thereby stimulating a number of works in short time \cite{NatCommunBPWeiJi,
BPNCJS4727, PhysRevB.89.235319, BPXWJ4458, PhysRevB.90.075434,
JPCLjz500409m, PhysRevB.89.245407, doi:10.1021/nl500935z,
apl-104-10-103106}, which also attracts much interest in once ignored bulk BP.
On the other hand, previous studies have promised that a strain engineering could modulate or even enhance the TE performance of relevant materials.
\cite{PhysRevB.87.125148, PhysRevB.86.184111, PhysRevB.84.165214,
aplSaeedSinghSchwingensBi2Se3, APL2013103031907}
Hence, it is worth to study the TE performance of bulk BP, especially under a
strain.

In this paper, by combining the density functional theory (DFT) first-principles
calculations and the semi-classical Boltzmann transport theory, we systematically investigated geometric, electronic, and TE properties of bulk BP under a strain. A number of unexpected and interesting results are disclosed. It is shown that the hinge-like structure of BP brings unusual mechanical response properties, such as the anisotropic Young's modulus and negative Poisson's ratio. The electronic structure of BP is observed being sensitive to a strain, as it can transform among phases of metal, direct and indirect semiconductors under a proper strain.
The maximal \emph{ZT} value of BP is found to be 0.72 along the $x$ direction at $800\,\mathrm{K}$, and can be enhanced to 0.87 by a proper strain, indicating that BP could be a potential $n$-type medium-high temperature TE material. The strain can be applied as an effective means to enhance TE properties. Such strain-induced enhancements are observed to occur often at the boundary of the direct-indirect band gap transition, which can be attributed to the increase of degeneracy of energy valleys at the transition point. The direct-indirect transition can act as a pointer of the multiple valleys and the enhancement of TE properties. On the basis of comparing BP with SnSe,  a family of potential good TE materials with the structure similar to BP and SnSe are predicted.


\begin{figure}
    \begin{minipage}{0.9\linewidth}
        \includegraphics[width=\textwidth]{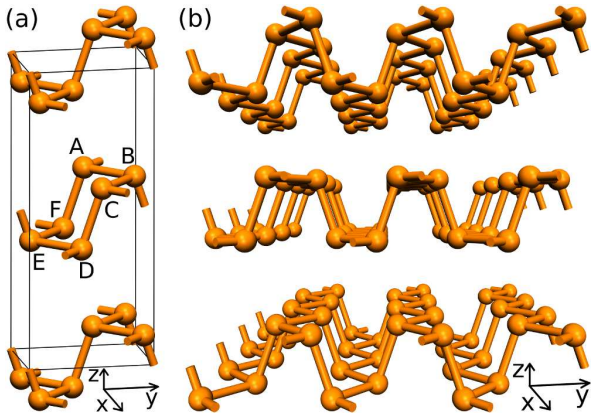}
    \end{minipage}
\caption{\label{fig:structure}
(a) A conventional cell and (b) a perspective side view of the crystal structure of black phosphorus. Within a hexagonal ring in one single layer, the upper three atoms are indicated as $A$, $B$ and $C$, while the nether three atoms are indicated as $D$, $E$ and $F$. The lattice parameters along $x$, $y$ and $z$ directions are defined as $a$, $b$ and $c$, respectively.
}
\end{figure}

\section*{Results}


\begin{figure}
    \begin{minipage}{\linewidth}
        \includegraphics[width=\textwidth]{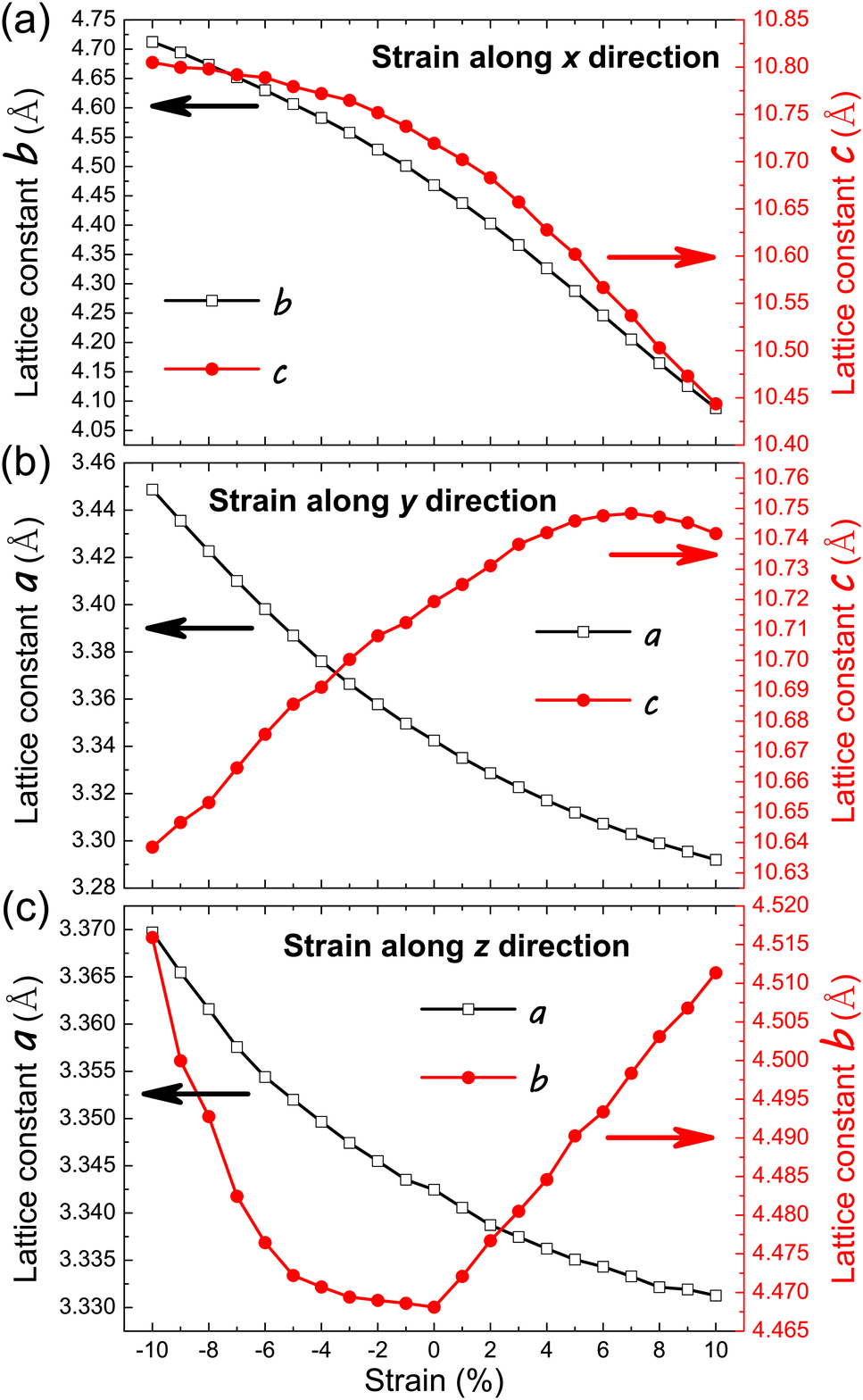}
    \end{minipage}
\caption{\label{fig:lattice}
Lattice parameter $l$ of bulk black phosphorus as function of the strain along $x$ (a), $y$ (b), and $z$ (c) directions. Strain is defined as $s =(l - l_0) / l_0$, where $l = a, b, c$ represent the lattice parameters along $x, y, z$ directions under strain, respectively, and $l_0 = a_0, b_0, c_0$ are the corresponding original lattice constants without strain.
}
\end{figure}

\textbf{Geometric structures under strain.} BP has a layered orthorhombic structure with space group \emph{Cmca} (No.\ 64), as shown in Fig.~\ref{fig:structure}, which is very similar to the structure of SnSe except its lower symmetry with space group \emph{Pnma} (No.\ 62). In BP,
P atoms within a single layer are covalently bonded with each other, forming
a puckered graphene-like hexagonal structure.
The pucker is so dramatic that a hinge-like structure along the $y$ direction is
formed and the in-plane anisotropy is obvious, which is different from that of
graphene.
Besides, it is also reported that the individual layers in BP are stacked together by \emph{van der Waals} Keesom forces\cite{jap-107-093718}, unlike London forces in graphite \cite{PhysRevB.73.153406}.
The optimized geometric structures of BP in our calculations are in good agreement with experimental and previous computational results \cite{jap-107-093718, Jpn.J.Appl.Phys.1984.23.15.Narita, Applied.Physics.A.1986.39.0947.Morita} (Supplemental Table S1).

The calculated elastic constants satisfy Born's mechanical stability criteria
and are in good agreement with experimental values and previous computational
results, implying the mechanical stability of the optimized structure of BP.
\cite{KHKNEMUSCBP52, YSFTEPBP53, PhysRevB.86.035105}
(Supplemental Table S2)
Now let us address the effect of the uniaxial strains along $x$, $y$ and $z$ directions on BP. The corresponding lattice parameters under strains are shown in Fig.~\ref{fig:lattice}. For clarity, the strain is defined as $s =(l - l_0) / l_0$, where $l = a, b, c$ are lattice parameters along the $x, y, z$ directions under strain, respectively, and $l_0 = a_0, b_0, c_0$ are the corresponding original lattice constants without strain. The positive (or negative) $s$ means a tensile (or compressive) strain, while $s=0$ corresponds the case without strain.
When the strain is along the $x$ direction (Fig.~\ref{fig:lattice}(a)), the lattice parameters $b$ and $c$ decrease with the increase of strain, giving rise to a normal positive Poisson's ratio, which is similar to the case in most materials.
When the strain is applied along the $y$ direction (Fig.~\ref{fig:lattice}(b)), however, the lattice parameter $a$ also decreases with the increase of strain, but the lattice parameter $c$ behaviors oppositely, resulting in a negative Poisson's ratio, which can be obtained as $-0.059$ by fitting $s = -\nu_1 t + \nu_2 t^2 + \nu_3 t^3$, where $s$ is the strain along the $z$ direction, $t$ is the variation of the lattice parameter along the $y$ direction, and $\nu_1$ could be regarded as the Poisson's ratio.
Surprisingly, when the strain is along the $z$ direction (Fig.~\ref{fig:lattice}(c)), the lattice
parameter $b$ first decreases and then increases with the increase of strain, where the minimum locates at the zero strain, implying an unexpected transition of Poisson's ratio from
positive ($0.012$) to negative ($-0.11$) that occurs at the critical point $s=0$. This observation suggests that whether the compressive or tensile strain is applied along the $z$ direction, BP will expand along the $y$ direction.
Such a phenomenon is sparse and could be explained mechanically.
When the strain along the $z$ direction is compressive, the expansion of BP
along the $y$ direction can be understood by considering the decrease of the
dihedral angle formed by the plane containing atoms marked by A, C and D
(abbreviated as plane ACD) and the plane ABC (see Fig.~1(a)).
Nevertheless, when the strain along the $z$ direction is tensile, the $\angle
ABC$ decreases, and the dihedral angle formed by plane ACD and ABC also
decreases because of the weakened \emph{van der Waals} interactions between
layers.
As a result, these factors together lead to the expansion of BP along the $y$
direction under tensile strain along the $z$ direction.
A detailed analysis on the layer thickness and the inter-layer distance of BP under strain can be found in Supplemental Fig.~S2, Fig.~S3 and related paragraphs.

\begin{figure}
    \begin{minipage}{\linewidth}
        \includegraphics[width=\textwidth]{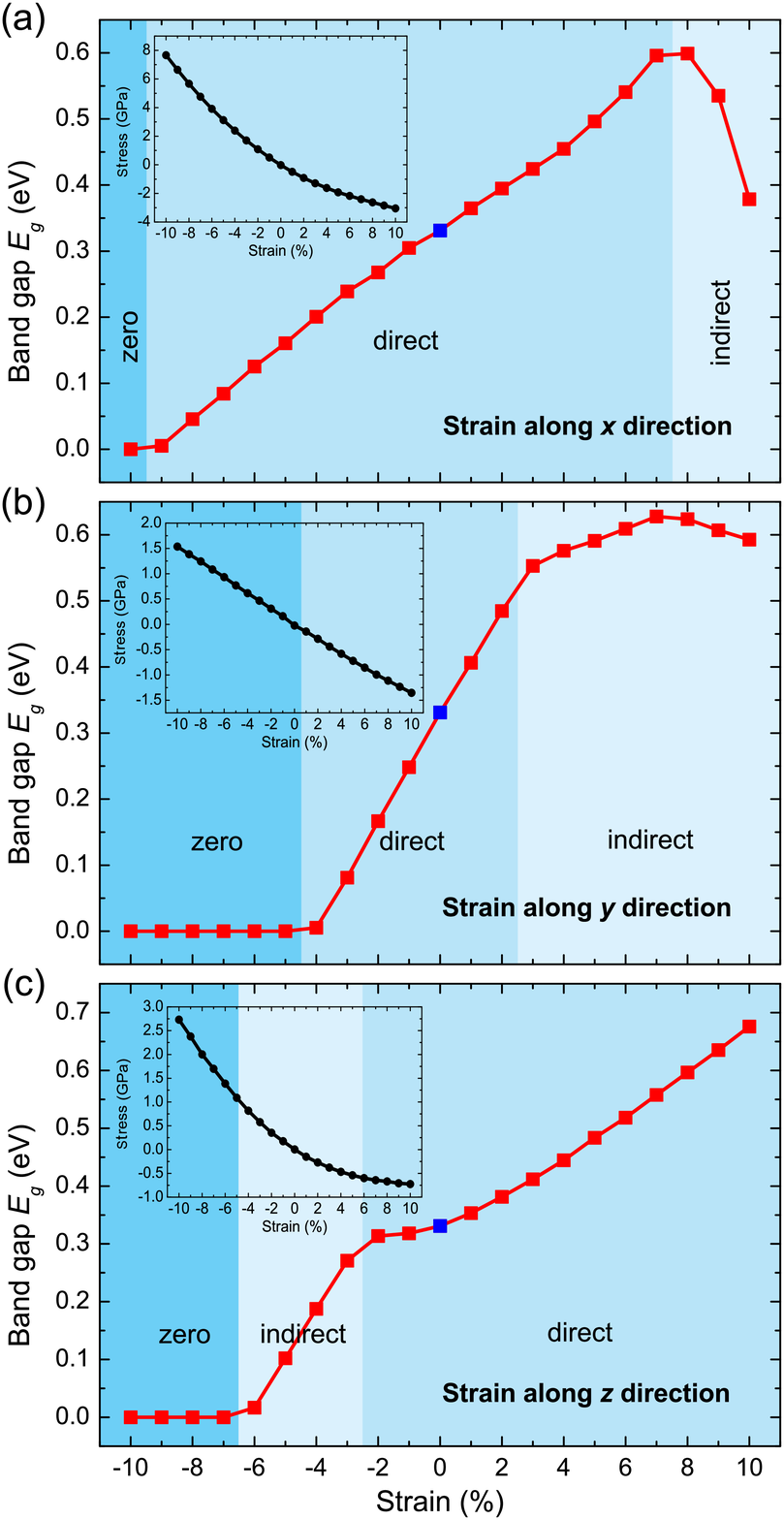}
    \end{minipage}
\caption{\label{fig:gap}
Band gap $E_g$ of bulk black phosphorus as function of the strain along $x$ (a), $y$ (b) and $z$ (c)
directions. The areas corresponding to direct, indirect and zero band gap are
indicated with different colors. The insets depict the relation between the stress and strain.
}
\end{figure}

Figure~\ref{fig:lattice}(c) also shows that the scale of $b$ is much larger than that of $a$, implying BP might be harder along the $x$ direction than along the $y$ direction. The Young's modulus along $x$, $y$ and $z$ directions can be evaluated to be $49.89\,\mathrm{GPa}$, $15.11\,\mathrm{GPa}$ and $15.68\,\mathrm{GPa}$ based on the slopes of the stress-strain curves plotted in the insets of Fig.~\ref{fig:gap}, revealing that BP is indeed much harder in $x$ direction than in $y$ or $z$ direction. The fact that Young's modulus along the $y$ direction is close to that along the $z$ is unexpected, as the interactions in BP along $y$ and $z$ directions are dominated by strong covalent bonds and weak \emph{van der Waals} forces, respectively.
The negative Poisson's ratio and the anisotropic Young's modulus both indicate that bulk BP has unusual mechanical responses, which is closely related to its layered hinge-like structure.



\textbf{Strain modulated electronic structure.} Bulk BP is a semiconductor.
Our calculations show that it has a direct band gap $E_g = 0.33\,\mathrm{eV}$, which agrees well with the experimental result \cite{jap.34.1853}. In experiment, the carrier mobilities of both $n$-type and $p$-type BP along the $y$ direction are much larger than those along the $x$ and $z$ directions, implying that the carrier transport is distinctly anisotropic, where the $y$ direction is most favored for conduction. It resembles SnSe that also has anisotropic conductivity and carrier mobilities \cite{ZLLSZYSHTGUCWCDK}.

The band structures of BP under different strains from $-10\pct$ to $10\pct$ along $x$, $y$ and $z$ directions are calculated (Supplemental Fig.~S4 and Fig.~S5), and the corresponding energy gaps $E_g$ are collected and visualized in Fig.~\ref{fig:gap}.
Interestingly, the energy gaps under strain could be either direct, indirect or even zero, which shows that strain can make bulk BP transform among the phases of metal, direct and indirect semiconductors, as indicated with different coloured areas in Fig.~\ref{fig:gap}. In general, the compressive strain may suppress the band gap, and finally closes the gap, while the tensile strain may enlarge the band gap \cite{C3CP44457K}. However, in bulk BP the situation varies under the strain along different directions.

When the strain is along the $x$ direction, the band gap of BP keeps direct and increases almost linearly when the strain increases from $-9\pct$ to $7\pct$ (Fig.~\ref{fig:gap} (a)). If the compressive strain is large enough, i.e., about $-10\pct$ (corresponding to an external pressure $7.67\,\mathrm{GPa}$) or more, the gap vanishes and BP turns into a metal. With increasing the strain, the direct band gap goes to a maximum (about $0.6\,\mathrm{eV}$) at $8\pct$  (corresponding to the tensile strength $2.63\,\mathrm{GPa}$), and then the band gap becomes indirect and declines quickly.

When the strain is along the $y$ direction, a similar behavior is observed. However, the range of direct band gap becomes narrow, in the interval of $-4\pct$ and $2\pct$ strain, in which the gap increases more steeply with the increase of strain (Fig.~\ref{fig:gap} (b)). The metal-semiconductor transition occurs at $-5\pct$ strain ($0.77\,\mathrm{GPa}$ external pressure), and the direct-indirect critical point is at $3\pct$ strain ($0.44\,\mathrm{GPa}$ tensile strength), which are nearly one order of magnitude lower than the pressure needed for the strain along the $x$ direction for such a transition. It is seen that BP is much softer along the $y$ direction than along the $x$ direction.

When the strain is along the $z$ direction, the band gap as a function of strain
displays a behavior distinctly different from those along the $x$ and $y$
directions (Fig.~\ref{fig:gap} (c)).  A tensile strain does not turn the band
gap of BP from direct into indirect within the range we considered (less than
$10\pct$), but makes the band gap increase almost linearly, while a small
compressive strain ($-3\pct$, $0.58\,\mathrm{GPa}$) turns the band gap from
direct to indirect. When the compressive strain reaches $-7\pct$
($1.7\,\mathrm{GPa}$), the band gap becomes zero and BP turns into a metal.
It is interesting to note that Narita \emph{et al}.\cite{Jpn.J.Appl.Phys.1984.23.15.Narita, Applied.Physics.A.1986.39.0947.Morita}  have observed experimentally a semiconducting to metallic transition of BP without phase change at the pressure around $1.70\,\mathrm{GPa}$ from
the resistivity-pressure measurements. Our calculated results are nicely consistent with this experimental observation.

\begin{figure}[t]
    \begin{minipage}{\linewidth}
        \includegraphics[width=\textwidth]{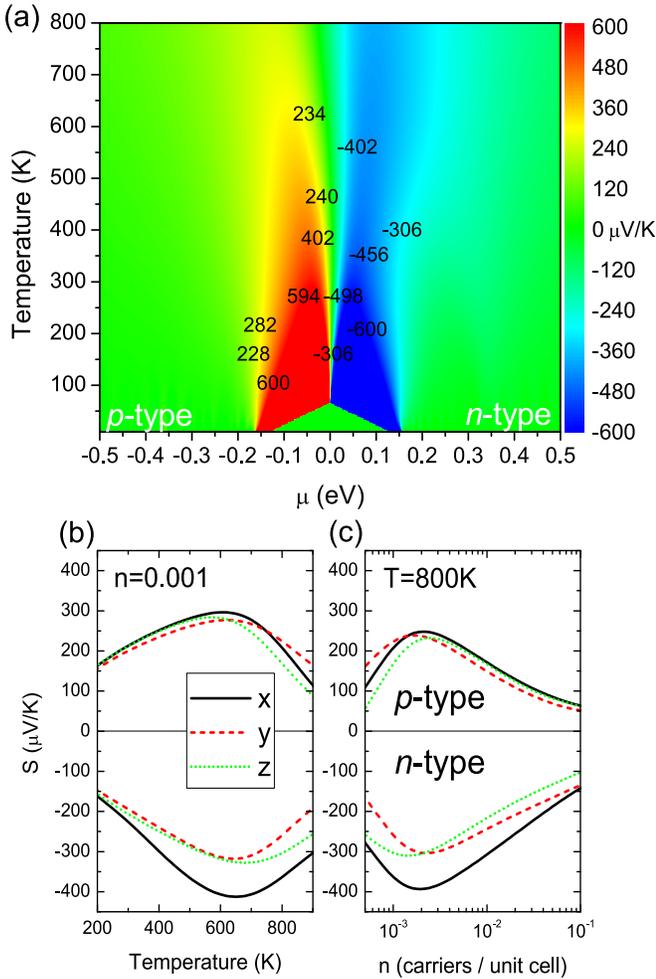}
    \end{minipage}
\caption{\label{fig:S}
(a) The contour plot of the thermopower ($S$) of bulk black phosphorus along the $x$ direction as a function of
chemical potential ($\mu$) and temperature.
(b) Thermopower ($S$) of bulk black phosphorus along $x$, $y$ and $z$ directions as a function of
temperature at the doping level of $0.001$ carriers per unit cell that corresponds
to a carrier concentration of $0.625\times 10^{19}\,\mathrm{cm^{-3}}$.
(c) Thermopower ($S$) of bulk black phosphorus along $x$, $y$ and $z$ directions as a function of doping
level at $800\,\mathrm{K}$.
}
\end{figure}
\begin{figure}[t]
    \begin{minipage}{\linewidth}
        \includegraphics[width=\textwidth]{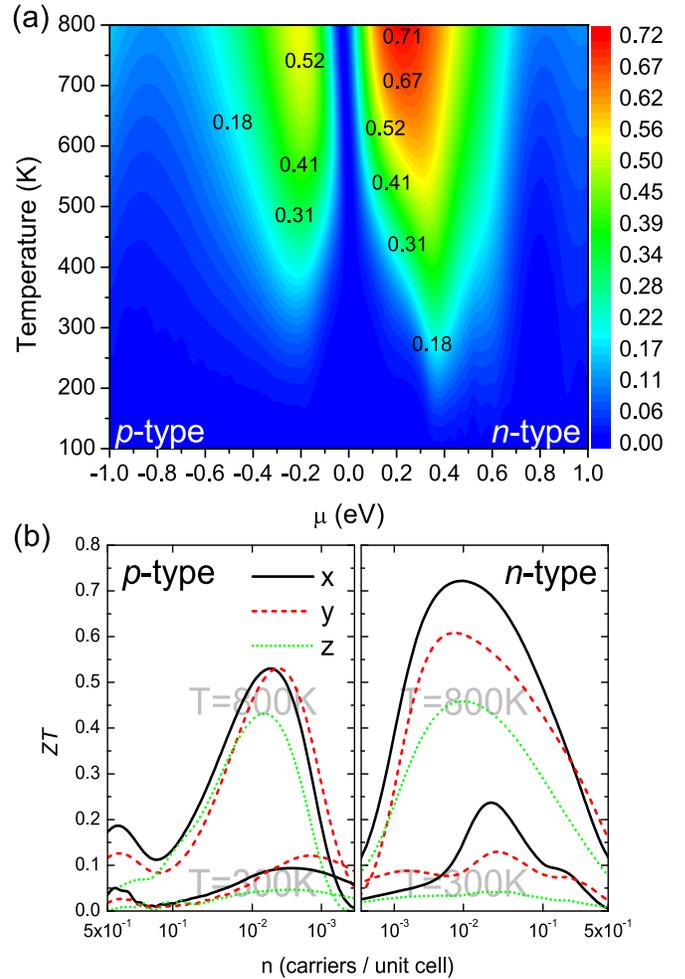}
    \end{minipage}
\caption{\label{fig:ZT}
(a) The figure of merit \emph{ZT} of bulk black phosphorus along the $x$ direction as a function of chemical
potential ($\mu$) and temperature ($T$).
(b) \emph{ZT} along $x$, $y$ and $z$ directions as a function of doping level at
$300\,\mathrm{K}$ and $800\,\mathrm{K}$ for both hole ($p$-type) and electron
($n$-type) doped bulk black phosphorus.
}
\end{figure}

\textbf{Thermoelectric properties without strain.} By means of the semi-classical Boltzmann transport theory, and by invoking the parameters from the calculated electronic structures and a few experimental values, the most important TE properties of bulk BP can be obtained. The thermopower $S$ and the figure of merit \emph{ZT} of BP as functions of temperature $T$ and chemical potential $\mu$ are presented in Fig.~\ref{fig:S} and Fig.~\ref{fig:ZT}.
Note that as the thermodynamic stable temperature and melting temperature of BP are $823\,\mathrm{K}$ and $883\,\mathrm{K}$, respectively
\cite{jap.34.1853, Applied.Physics.A.1986.39.0947.Morita}, we only consider the
TE properties of BP up to $800\,\mathrm{K}$.

The thermopower $S$ of BP along the $x$ direction is shown in
Fig.~\ref{fig:S}(a), which gives us an overall picture of the thermopower as a function of the temperature $T$ and chemical potential $\mu$, as it is qualitatively similar along the $y$ and $z$ directions (Supplemental Fig.~S6). The absolute value of $S$ indicates the actual power of TE conversion. A negative $\mu$ corresponds to hole ($p$-type) doping and gives a positive $S$, while a positive $\mu$ corresponds to electron ($n$-type) doping and gives a negative $S$.
Obviously, a large $S$ can only be obtained at a low doping concentration with $\mu$ between $-0.15\,\mathrm{eV}$ and $0.15\,\mathrm{eV}$; furthermore, a large $S$ more than $600\,\mathrm{\mu V/K}$ could be achieved at the temperature below $400\,\mathrm{K}$.
It is seen that $S$ of BP is asymmetric for $p$-type and $n$-type doping, and the latter is more preferred when temperature is higher than 400 K, which may be owing to the asymmetry of valence and conduction bands of BP.
The thermopower $S$
along three directions as function of $T$ at the fixed doping level of
$0.001\,\mathrm{carriers/unit\ cell}$ (corresponding to the carrier concentration $0.625\times 10^{19}\,\mathrm{cm^{-3}}$) and as function of doping level at a fixed temperature
$800\,\mathrm{K}$ are shown in Fig.~\ref{fig:S}(b) and Fig.~\ref{fig:S}(c), respectively.
$S$ is anisotropic, which is the largest along the $x$ direction for $n$-type doping.
When temperature increases, $S$ increases until the temperature reaches about $600 \sim 700\,\mathrm{K}$,
and the maximal $S$ is about $400\,\mathrm{\mu V/K}$ along the $x$ direction for $n$-type doping, while it is only $300\,\mathrm{\mu V/K}$ for other situations; $S$ decreases quickly when temperature over $600\,\mathrm{K}$ and continues to decrease, which may be caused by the bipolar conduction.
From Fig.~\ref{fig:S}(c), the best carrier concentration can be estimated to be $0.01 \sim 0.001\,\mathrm{carriers/unit\ cell}$. The thermopower $S$
along the $x$, $y$ and $z$ directions as function of doping level at fixed temperature
$300\,\mathrm{K}$ and $500\,\mathrm{K}$ are shown in supplemental Fig.~S8, which indicates that the anisotropic thermopower is more noticeable at high temperature.

The efficiency of thermoelectric conversion is characterized by the dimensionless figure of merit \emph{ZT}.
The value of \emph{ZT} of bulk BP along the $x$ direction as function of temperature $T$ and chemical potential $\mu$ is shown in Fig.~\ref{fig:ZT}(a), and those along the $y$ and $z$ directions are presented in supplemental Fig.~S7.
It can be seen that \emph{ZT} shows much more notable asymmetry for $p$-type and $n$-type doping.
The largest \emph{ZT} (0.72) is observed along the $x$ direction at temperature $800\,\mathrm{K}$ for an $n$-type doping concentration of $6.0\times 10^{19}\,\mathrm{cm^{-3}}$, while for the $p$-type doping, the maximal \emph{ZT} (0.53) is obtained along the $y$ direction at $800\,\mathrm{K}$ for a hole doping concentration of $2.72\times 10^{19}\,\mathrm{cm^{-3}}$.
Thus the properly doped BP should be a potential TE material working at medium-high temperature.
\emph{ZT} along the $x$, $y$ and $z$ directions as function of doping level at $300\,\mathrm{K}$ and $800\,\mathrm{K}$ are given in Fig.~\ref{fig:ZT}(b). For the $p$-type doping, the maximal \emph{ZT} along the $x$ and $y$ directions are nearly the same, while that along the $z$ direction is much smaller. For the $n$-type doping, the maximal \emph{ZT} along the $x$ direction is distinctly larger than those along the $y$ and $z$ directions, which is mainly attributed to the outstanding maximal $S$ (about $400\,\mathrm{\mu V/K}$) along the $x$ direction in this case. It is observed that the maximal \emph{ZT} at $800\,\mathrm{K}$ is much larger than that at $300\,\mathrm{K}$ with the corresponding doping concentration at $800\,\mathrm{K}$ less than that at $300\,\mathrm{K}$, which is caused by the increase of the electrical conductivity and the decrease of the lattice thermal conductivity with the increase of temperature. The maximal \emph{ZT} along the $x$, $y$ and $z$ directions as function of temperature are shown in supplemental Fig.~S9, which manifests that the bulk BP is indeed a medium-high temperature TE material.

\begin{figure*}[t]
    \begin{minipage}{0.9\linewidth}
        \includegraphics[width=\textwidth]{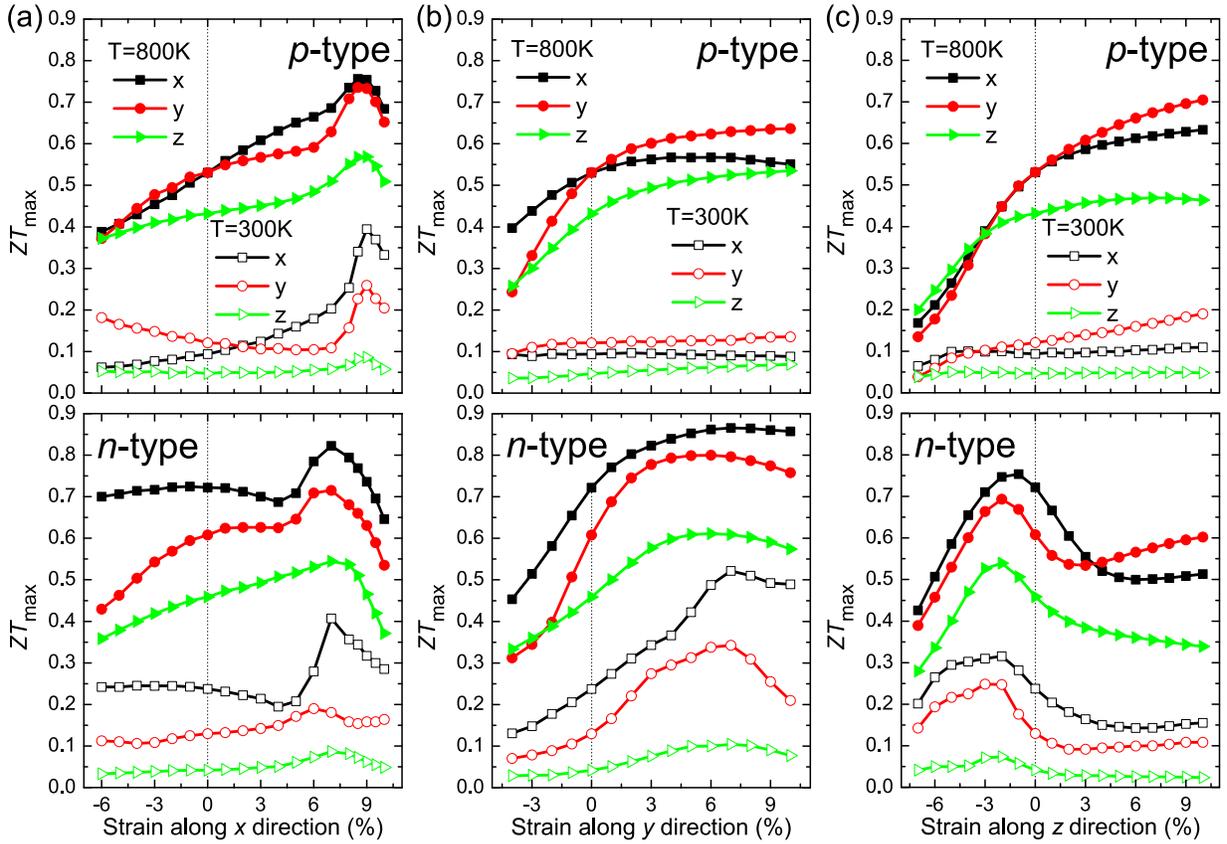}
    \end{minipage}
\caption{\label{fig:ZT-strain}
Maximal \emph{ZT} values along (a) $x$, (b) $y$ and (c)
$z$ directions for the hole ($p$-type) and electron ($n$-type) doped black
phosphorus as function of the strain applied along $x$, $y$ and $z$ directions at $300\,\mathrm{K}$ and $800\,\mathrm{K}$.
}
\end{figure*}

\textbf{Strain-modulated figure of merit \emph{ZT}.}
For the strains exerted along the $x$, $y$ and $z$ directions, \emph{ZT} of bulk BP as a function of temperature $T$ and chemical potential $\mu$ is calculated, and the corresponding maximal values at $300\,\mathrm{K}$ and $800\,\mathrm{K}$ are extracted, as shown in Fig.~\ref{fig:ZT-strain}. It should be noticed that, as the TE properties of BP are anisotropic, for the strain applied along every specific direction, \emph{ZT} of BP along the three directions are different.

Fig.~\ref{fig:ZT-strain}(a) exposes the effect of the strain along the $x$ direction on \emph{ZT} of $p$-type and $n$-type doped bulk BP. For the $p$-type doping, the tensile strain along the $x$ direction brings an obvious enhancement on $ZT$ at $800\,\mathrm{K}$. The maximal \emph{ZT} increases from about 0.39 to 0.76 ($x$ direction), 0.38 to 0.74 ($y$ direction) and 0.38 to 0.58 ($z$ direction) with the increase of the strain from $-6\pct$ to $8\pct$, and then decreases quickly when the strain is larger than $8\pct$, giving rise to a peak of \emph{ZT} at tensile strain $7\sim 8\pct$.
At $300\,\mathrm{K}$, the strain along the $x$ direction also leads to peaks of \emph{ZT} at $8\pct$ tensile strain, but the corresponding largest \emph{ZT} along the $x$ direction is only about 0.40, and is even smaller than those along the $y$ and $z$ directions.
For the $n$-type doping, when the strain is between $4\pct$ and $10\pct$, \emph{ZT} along the $x$, $y$, and $z$ directions all reveal the peaks at about $6\pct$ tensile strain, and the corresponding largest \emph{ZT} values are 0.82, 0.71 and 0.55, respectively.
Recall that the largest \emph{ZT} (0.72) of BP without a strain is observed along the $x$ direction. Hence, the strain along the $x$ direction could indeed enhance \emph{ZT} of BP along the $x$ direction for the $n$-type doping, with $14\pct$ larger than the value without a strain.
Besides, the largest \emph{ZT} for the $n$-type doped BP under the strain along the $x$ direction at $300\,\mathrm{K}$ is only about 0.40, which is the same as that of the $p$-type doping, implying that the TE properties of BP under strain are relatively poor at $300\,\mathrm{K}$.

The effect of the strain along the $y$ direction on \emph{ZT} of the $p$-type and $n$-type doped BP is disclosed in Fig.~\ref{fig:ZT-strain}(b). Although the \emph{ZT} value at $800\,\mathrm{K}$ monotonically increases with the increase of the strain, but the maximal \emph{ZT} is only 0.64 (along the $y$ direction), which occurs at $10\pct$ tensile strain.
At $300\,\mathrm{K}$, the \emph{ZT} values are nearly unaffected by the strain.
Obviously, the enhancement of $ZT$ for the $p$-type doped BP by the strain along the $y$ direction is not as much as that by the strain along the $x$ direction.
Contrarily, the enhancement of \emph{ZT} for the $n$-type doped BP by the strain along the $y$ direction is dramatic. With increasing the strain, \emph{ZT} at $800\,\mathrm{K}$ increases monotonically and rapidly, and the maximal \emph{ZT} values are 0.87 ($x$ direction), 0.80 ($y$ direction) and 0.61 ($z$ direction) under $5\sim7\pct$ tensile strain,  where \emph{ZT} values along the $x$ and $y$ directions are about $20\pct$ and $11\pct$ larger than the maximal \emph{ZT} value without a strain, respectively.
Besides, \emph{ZT} along the $x$, $y$, and $z$ directions at $300\,\mathrm{K}$ show peaks at about $7\pct$ tensile strain, and the corresponding maximal \emph{ZT} value along the $x$ direction reaches about 0.52, which is more than twice of the value (0.24) without a strain, and is also larger than that with the strain along the $x$ direction.

The upper panel of Fig.~\ref{fig:ZT-strain}(c) shows the effect of the strain along the $z$ direction on \emph{ZT} of the $p$-type doped BP, which is rather similar to that of the case with a strain along the $y$ direction (the upper panel of Fig.~\ref{fig:ZT-strain}(b)), but the increase of \emph{ZT} value with increasing the strain is more steeper. The maximal \emph{ZT} (0.70) at $800\,\mathrm{K}$ is found along the $y$ direction. For the $n$-type doping (the lower panel of Fig.~\ref{fig:ZT-strain}(c)), the effect of the strain along the $z$ direction on \emph{ZT} is rather different. The tensile strain does not enhance but weakens \emph{ZT}. The peaks of \emph{ZT} emerge at $-1\pct$ ($x$ direction) or $-2\pct$ (other directions) strain. At $800\,\mathrm{K}$, the maximal \emph{ZT} values are 0.75 ($x$ direction), 0.70 ($y$ direction) and 0.46 ($z$ direction), showing the enhancement is somewhat weak.

It is interesting to compare Fig.~\ref{fig:ZT-strain} with Fig.~\ref{fig:gap}. We discover that some of the strain-induced \emph{ZT} enhancements happen to appear at the boundary of the strain-induced direct-indirect transition of the energy gaps. For instance, a direct-indirect transition occurs at about $7\sim8\pct$ tensile strain along the $x$ direction (Fig.~\ref{fig:gap}(a)), where the maximal \emph{ZT} (indicated by peaks) along the three directions are obtained for both $p$-type and $n$-type doped BP (Fig.~\ref{fig:ZT-strain}(a)); in addition, for the $n$-type doped BP, the peaks of \emph{ZT} emerging at $-1\sim-2\pct$ compressive strain along the $z$ direction (the lower panel of Fig.~\ref{fig:ZT-strain}(c)) are related to the direct-indirect transition of the energy gaps.

From above results we could find that the effect of the strain on the TE performance of BP along the $x$ direction is more effective for the $p$-type doping than for the $n$-type doping, but the strain along the $y$ direction is more effective for the $n$-type doping than for the $p$-type doping.

\section*{Discussion}

BP has a layered hexagonal structure similar to the high-$ZT$ material SnSe. Each layer of BP can be viewed similar to the structure of well-known graphene but significantly puckered. It reminds us that the single layer of BP should be another elemental two-dimensional material besides graphene and silicene. However, the in-plane anisotropy of the geometrical structure of BP could be the key fcator for the dominant difference between BP sheet and graphene, which leads to the unexpected mechanical, electronic and TE properties as we presented above.

BP has parallel "grooves" and "roads" alternatively assembled to form the hinge-like structure, where the "grooves" and "roads" are composed of zigzag covalent bond of P atoms, as shown in Fig.~\ref{fig:structure}. Such a structure reveals that the mechanical strength of BP could be much weaker than that of graphene, and BP is harder along the $x$ direction than along the $y$ direction. This is confirmed by the anisotropic Young's modulus along the $x$ ($49.89\,\mathrm{GPa}$) and $y$ ($15.11\,\mathrm{GPa}$) directions obtained by the DFT calculations.
It is interesting to find that the Young's modulus along the $y$ direction is very close to that along the $z$ direction ($15.68\,\mathrm{GPa}$). This result is surprising that BP along the $y$ direction appears to be as soft as that along the $z$ direction, because in the former case every P atom is connected through covalent bonds while in the latter case the atoms between layers are bonded the \emph{van der Waals}. This anisotropic property of BP could be a result of the hinge-like assembling structure.
In addition, when the strains are applied along $y$ and $z$ directions, the negative Poisson's ratio of BP is observed, which can also be attributed to the hinge-like structure that were shown to be effective to form auxetic materials \cite{Almgren, Valant, Shufrin, Pasternak}. In this sense, SnSe and other similar materials with hinge-like layered structures might have anisotropic mechanical properties and negative Poisson's ratio, and could be potential auxetic materials.

The electronic structure of BP is sensitive to the strain. It is uncovered that the strain along the $x$, $y$ and $z$ directions can give rise to the transitions among metal, direct and indirect semiconductors.
The compressive pressure needed for the metal-semiconductor transition in BP is $7.67\,\mathrm{GPa}$ ($x$ direction), $0.77\,\mathrm{GPa}$ ($y$ direction), and $1.70\,\mathrm{GPa}$ ($z$ direction), illustrating that
the conduction property of the BP can be readily manipulated by a moderate compressive or tensile strain, which may have potential applications in nanoelectronics.

The thermopower ($S$) and figure of merit (\emph{ZT}) of BP in the absence of the strain are nonlinear with temperature $T$ and chemical potential $\mu$ (Fig.~\ref{fig:S}(a) and Fig.~\ref{fig:ZT}(a)). The best combination of $T$ and $\mu$ for a better \emph{ZT} can be extracted. However, the $T$-$\mu$ area for large \emph{ZT} does not overlap with that for large $S$. This fact indicates that a large thermopower will not certainly lead to a high
\emph{ZT} value; on the contrary, a moderate thermopower combined with a
suitable electrical and thermal conductivity may eventually result in a high
\emph{ZT} value. The similar phenomenon was also observed in experiments
\cite{ZLLSZYSHTGUCWCDK}.
Although thermopower is an important indicator when searching for TE materials with excellent performance, the attention should also be paid to the electrical and thermal conductivity of the materials.

The calculated maximal \emph{ZT} (0.72) of BP in the absence of a strain is distinctly smaller than that (2.6) of SnSe, which can be understood in this way. Although BP has an anisotropic layered structure similar to SnSe, the symmetry of BP is higher than that of SnSe; BP is an elemental phosphorus, while SnSe comprises of two different types of atoms; thus the bonding and lattice anharmony of SnSe are much stronger than those of BP. Also, Sn and Se atoms are much heavier than P atom.
These factors would lead to the lattice thermal conductivity of BP is much larger than that of SnSe, giving the \emph{ZT} value of BP much lower than SnSe. If we use two different types of heavier atoms to replace the phosphorus atoms in BP, getting compounds such as GeSe, SnO, SnS, etc., it will lead to a family of potential good TE materials with structures similar to BP and SnSe.

The strain can enhance the TE performance of BP, for instance, the \emph{ZT} can be promoted from 0.53 to 0.76 ($p$-type) and from 0.72 to 0.87 ($n$-type) at $800\,\mathrm{K}$.
We mentioned that some of the strain-induced \emph{ZT} enhancements happen to emerge at the boundary of the direct-indirect gap transition of the energy bands. In fact, it is not accidental, which is closely related to the strain-modulated electronic structure (See supplementary Fig.~S5).
For BP under no strain, the valence band maximum (VBM) and conduction band minimum (CBM) sit at $\Gamma$, leaving a $0.33\,\mathrm{eV}$ direct band gap. If a strain is applied, e.g., along the $x$ direction, one may observe that when the tensile strain increases, the CBM at $\Gamma$ ascends and the conduction band between $Y$ and $\Gamma$ points descends concurrently, and align in energy at the strain of $7\sim 8\pct$; eventually the conduction band between $Y$ and $\Gamma$ points becomes the new CBM and induces an indirect band gap. In other words, in the direct-indirect gap transition under a strain, two or more conduction bands (valence bands) will converge and then depart, i.e., the degeneracy of energy valleys is increased at the transition point. Note that the "multiple valleys" are reported as one of the most important factors in the improvement of TE properties \cite{Pei}. Thus, the connection between the direct-indirect band gap transition and the \emph{ZT} value enhancement is  clear.
Although the "multiple valleys" are not always accompanied by a direct-indirect gap transition,  the latter can increase the degeneracy of energy valleys. From this observation, we find that the direct-indirect gap transition can act as a pointer of the multiple valleys and the enhancement of TE properties regardless of the reason for such a transition, which may be useful for searching for the good TE materials.

In summary, we systematically investigated the geometric, electronic, and TE
properties of bulk BP under strain, and found that BP could be as a potential TE material.
It is revealed that bulk BP possesses many interesting and unexpected geometric
properties, such as hinge-like structures, anisotropic Young's modulus, and negative Poisson's ratio. BP can transit among metal, direct and indirect semiconductors under a strain, showing a sensitive electronic structure that could lead to diverse modulation possibilities.
The TE performance of BP is anisotropic, where the maximal \emph{ZT} value is found to be 0.72 along the $x$ direction at $800\,\mathrm{K}$, indicating BP may be a potential $n$-type medium-high temperature TE material. It is disclosed that the strain is able to modulate \emph{ZT} of BP effectively, and the \emph{ZT} value is enhanced to 0.87 at a tensile strain along the $y$ direction.
Interestingly, most enhancements of the TE performance of BP happen to emerge at the boundary of the direct-indirect transition of the energy bands, which may be attributed to the increase of the degeneracy of energy valleys at the transition point. The direct-indirect gap transition can play a role as an indicator of the multiple valleys and the enhancement of TE properties. By comparing the structure of BP with SnSe, a family of potential TE materials with structures similar to BP and SnSe are suggested. This present study not only presents various unexpected properties of BP, but also gives hints that could be applied to enhance the TE effect of a material.

\section*{Methods}

The first-principles DFT calculations are performed using the projector augmented
wave method \cite{PhysRevB.59.1758} and the generalized gradient approximation
(GGA) of Perdew-Burke-Ernzerhof (PBE)\cite{PhysRevLett.77.3865} for the
exchange-correlation potential as implemented in the Vienna \emph{ab-initio}
simulation package (\texttt{\textsc{vasp}}) code\cite{PhysRevB.54.11169}.
The kinetic energy cutoff of wave functions is $700\,\mathrm{eV}$, and a
Monkhorst-Pack \cite{PhysRevB.13.5188} $k$-mesh of $10\times 8\times 4$ is adopted
to sample the inreducible Brillouin zone (IBZ), with the energy convergence
threshold set as $10^{-8}\,\mathrm{eV}$.
\emph{Van der Waals} interaction is taken into account at the vdW-DF level with
optB88 for exchange functional (optB88-vdW)\cite{PhysRevB.83.195131,
0953-8984-22-2-022201}.
Both the cell shape and volume are fully optimized and all atoms are allowed to
relax until the maximal Hellmann-Feynman force acting on each atom is less than
$0.001\,\textrm{eV/\AA}$.
As the exact band gap is important for the accurate prediction of the
TE transport properties\cite{jcp-1.3666851}, the electronic
structures are calculated at modified Becke-Johnson
(mBJ) \cite{PhysRevLett.102.226401} level.

Once the full electronic band structures are obtained with dense enough k-points
sampled in the first Brillouin zone, several TE properties, such as thermopower ($S$), electrical
conductivity ($\sigma$) and electrical thermal conductivity (${\kappa}_e$) can
be derived using the semi-classical Boltzmann transport theory as implemented in the
\texttt{\textsc{boltztrap}} code\cite{Madsen200667}, which has been shown to
provide a good description of TE properties in a variety of
TE materials.\cite{PhysRevB.82.035204, PhysRevB.83.115110,
PhysRevB.84.125207, PhysRevB.85.165149, PhysRevB.86.155204}
Note that several approximations are involved in the above method.
As the band structures are calculated at zero temperature, the broadening of
Fermi distribution has been utilized to introduce the effect of
temperature\cite{Madsen200667}.
Rigid band approximation \cite{aplmaterials-011101,
PhysRevB.85.165149} is employed that doping could be simulated by simply
shifting up ($n$-type) or down ($p$-type) the chemical potential based on the
undoped band structures, which means that only low level doping could be
treated properly.
Besides, the constant scattering time ($\tau$) approximation is adopted to ignore the $k$ and $E$-dependence of the scattering time $\tau$, but it still can be varying for different materials or different directions in the same material.

In fact, the directly obtained ($\sigma$) and ${\kappa}_e$ from \texttt{\textsc{boltztrap}} code are expressed as the ratios of electrical conductivity, electrical thermal conductivity to the scattering time $\tau$, i.e., $\sigma/\tau$ and ${\kappa}_e/\tau$, the figure of merit\emph{ZT} could be rewritten as
$$ZT = \frac{S^2\sigma T}{\kappa_e+\kappa_{ph}} = \frac{S^2 \frac{\sigma}{\tau} T}{\frac{\kappa_e}{\tau}+\frac{\kappa_{ph}}{\tau}}$$
Hence, to obtain \emph{ZT}, we still need to know $\kappa_{ph}/\tau$.

For such an anisotropic material BP,  it has been reported
\cite{JPSJ.52.2148} that there exists a contradiction of the ratio of mobility
to reciprocal effective mass between the experiment and the results obtained with a
simple consideration that the scattering times are comparable among the
crystal directions. Hence the anisotropy of $\tau$ is necessary to be considered.
From the semiconductor theory \cite{TCTCEM04}, the carrier's
scattering time $\tau$ could be derived from the relation $\mu=e\tau/m^{*}$,
where $\mu$ is the carrier mobility, $m^{*}$ is the effective mass of carrier, and $e$
is the elementary charge.
With $\mu$ and $m^{*}$ along different lattice directions for both electron
($n$-type) and hole ($p$-type) doped BP extracted from the experiment
\cite{Applied.Physics.A.1986.39.0947.Morita}, the anisotropic $\tau$ for BP can
be evaluated (See supplemental Table S3).
For the lattice thermal conductivity ($\kappa_{ph}$), which could be derived in
principle by considering the phonon dispersions and two-phonon or even higher
order scattering mechanisms,  we here take into account the influence
of $\kappa_{ph}$ to the TE performance of BP through a more practical approach:
consider an inverse temperature dependence of the lattice thermal
conductivity ($\kappa_{ph}$) that  exists in a large number of materials \cite{PhysRevB.82.035204,
PhysRevB.86.155204, lina2013}, we could simply assume that $\kappa_{ph}
\propto 1/T$ also for BP. With the lattice thermal conductivity $\kappa_{ph}(300\,\mathrm{K}) =
12.1\,\mathrm{W/mK}$ obtained from the experiment \cite{PhysRev.139.A507}, the
$\kappa_{ph}(T)$ for other higher temperatures could be derived using the
reciprocal relation of $\kappa_{ph}$ to $T$. Thus, the $\kappa_{ph}/\tau$ can be calculated.



\section*{Acknowledgments}
\noindent
The authors thank Prof.\ Ming Hu of RWTH, Prof.\ Wei Ji of RUC and Prof.\ Zhen-Gang Zhu of UCAS for their helpful discussions. All calculations are performed on Nebulae (DAWN6000) in National Supercomputing Center in Shenzhen and MagicCube (DAWN5000A) in Shanghai Supercomputer Center, China. This work is supported in part by the NSFC (Grant No.\ 11004239), the MOST (Grant No.\ 2012CB932901 and No.\ 2013CB933401) of China, and the fund from CAS.

\section*{Author contributions}
\noindent
Q.B.Y., Q.R.Z. and G.Z.Q. conceived and designed the research.
G.Z.Q., Z.Z.Q. and S.Y.Y. carried out the calculations.
G.S., Q.B.Y. and G.Z.Q. co-wrote the paper with contribution from Z.Z.Q and H.J.C..
All the authors participated in the discussions and reviewed the manuscript.
G.S. supervised the whole project.

\section*{Additional information}

\noindent\textbf{Supplementary information}
accompanies this paper at http://www.nature.com/scientificreports

\noindent\textbf{Competing financial interests:}
The authors declare no competing financial interests.

\noindent\textbf{How to cite this article:}
Qin, G. \emph{et al.}~Hinge-like structure induced unusual properties of black
phosphorus and new strategies to improve the thermoelectric performance. 
\emph{Sci.~Rep.~}\textbf{4,} 6946; DOI:10.1038/srep06946 (2014).

\end{document}